\documentclass[onecolumn,showpacs,10pt]{revtex4}

\usepackage{graphicx}% Include figure files
\usepackage{dcolumn}% Align table columns on decimal point
\usepackage{bm}% bold math
\usepackage{color}
\usepackage{amssymb,amsmath}

\input epsf

\begin{document}

\title{\Large Accretions of Dark Matter and Dark Energy onto ($n+2$)-dimensional Schwarzschild
Black Hole and Morris-Thorne Wormhole}

\author{\bf Ujjal Debnath\footnote{ujjaldebnath@gmail.com ,
ujjal@iucaa.ernet.in}}

\affiliation{Department of Mathematics, Indian Institute of
Engineering Science and Technology, Shibpur, Howrah-711 103,
India.\\}

\date{\today}

\begin{abstract}
In this work, we have studied accretion of the dark matter and
dark energy onto of $(n+2)$-dimensional Schwarzschild black hole
and Morris-Thorne wormhole. The mass and the rate of change of
mass for $(n+2)$-dimensional Schwarzschild black hole and
Morris-Thorne wormhole have been found. We have assumed some
candidates of dark energy like holographic dark energy, new
agegraphic dark energy, quintessence, tachyon, DBI-essence, etc.
The black hole mass and the wormhole mass have been calculated in
term of redshift when dark matter and above types of dark energies
accrete onto them separately. We have shown that the black hole
mass increases and wormhole mass decreases for holographic dark
energy, new agegraphic dark energy, quintessence, tachyon
accretion and the slope of increasing/decreasing of mass
sensitively depends on the dimension. But for DBI-essence
accretion, the black hole mass first increases and then decreases
and the wormhole mass first decreases and then increases and the
slope of increasing/decreasing of mass not sensitively depends on
the dimension.
\end{abstract}


\maketitle

\section{Introduction}

Recent observations of type Ia supernovae indicate that the
expansion of the Universe is accelerating rather than slowing down
\cite{{Bachall},{Perlmutter1},{Perlmutter2},{Riess}}. These
results, when combined with cosmic microwave background (CMB)
observations of a peak in the angular power spectrum on degree
scales \cite{Bern,Lange,Babli}, strongly suggest that the Universe
is spatially flat with $\sim$ 1/3 of the critical energy density
being in non-relativistic matter and $\sim$ 2/3 in a smooth
component with large negative pressure. This acceleration is
caused by some unknown matter is known as ``dark energy'' (DE)
\cite{Briddle,Spergel,Peebles,Cald}. Recent WMAP \cite{Bennett}
and Chandra X-ray Observations \cite{Allen} strongly indicate that
our universe is undergoing an accelerating phase. The most
appealing and simplest candidate for DE is the cosmological
constant $\Lambda$ which is characterized by the equation of state
$p=w\rho$ with $w=-1$. Many other theoretical models have been
proposed to explain the accelerated expansion of the universe.
Another candidate of dark energy is quintessence satisfying
$-1<w<-1/3$ \cite{Peebles,Cald}. When $w<-1$, it is known as
phantom energy \cite{Alam2} which has the negative kinetic energy.
Recently many cosmological models have been constructed by
introducing dark energies such as quintessence
\cite{Peebles,Cald}, DBI-essence \cite{Gum,Mart}, Tachyon
\cite{Sen}, holographic dark energy \cite{Li,Hsu,Cohen}, new
agegraphic dark energy \cite{Cai01,wei01,wei02}, etc.\\

In Newtonian theory, the problem of accretion of matter onto the
compact object was first formulated by Bondi \cite{Bondi}. Michel
\cite{Michel} has formulated the accretion process of steady-state
spherical symmetric flow of matter into or out of a condensed
object. The accretion of phantom dark energy onto a static
Schwarzschild black hole was first formulated by Babichev et al
\cite{Babichev,Babichev1} using Michel's process and established
that static Schwarzschild black hole mass will gradually decrease
to zero near the big rip singularity. Recently, Jamil \cite{Jamil}
has investigated accretion of phantom like modified variable
Chaplygin gas onto Schwarzschild black hole. Also the accretion of
dark energy onto the more general Kerr-Newman black hole was
studied by Madrid et al \cite{Pedro} and Bhadra et al
\cite{Bhadra}. Several authors \cite{Nayak,Lima,Sharif1,Sun,Kim,
Rod,Abhas,Mar1,Abha} have discussed the accretions of various
components of dark energy onto several types of black holes. Now
there is a lot of interest of the investigation of dark energy
accretion onto static wormhole \cite{Diaz1,Far,Diaz2}.
Gonz$\acute{a}$lez-D$\acute{i}$az \cite{Diaz3} has discussed the
phantom energy accretion onto wormhole. Madrid et al \cite{Mad}
and Mart$\acute{i}$n-Moruno \cite{Mor} have analyzed a general
formalism for the accretion of dark energy onto astronomical
objects, black holes and wormholes. Subsequently, the in
accelerating universe, the dark energy accretion onto wormhole has
been discussed in \cite{Diaz4,Diaz5}.\\

Recently, there has been a growing interest to study the accretion
of higher dimensional black hole (BH). Interest in the BTZ black
hole has recently heightened with the discovery that the
thermodynamics of higher dimensional black holes \cite{Kim1}.
Also, non-static charged BTZ like black holes in
$(n+1)$-dimensions have been considered by Ghosh et al
\cite{Ghosh}. John et al \cite{John} examined the steady-state
spherically symmetric accretion of relativistic fluids (like
polytropic equation of state) onto a higher dimensional
Schwarzschild black hole. Also charged BTZ-like black holes in
higher dimensions have been studied by Hendi \cite{Hendi}. By the
motivations of above works, we shall assume the accretions of dark
matter and dark energy onto ($n+2$)-dimensional Schwarzschild
black hole and Morris-Thorne wormhole. The nature of masses of
black hole and wormhole will be investigated during various types
of dark energies like holographic dark energy, new agegraphic dark
energy, quintessence, tachyon, DBI-essence, etc. Finally we
present
the conclusions of the whole work.\\

\section{Accretions of dark  matter and energy onto Schwarzschild black hole and Morris-Thorne Wormhole}

Let us consider $(n+2)$-dimensional spherically symmetrical
accretion of the dark energy onto the black hole. We consider a
Schwarzschild black hole (static) of mass $M$ which is
gravitationally isolated (in geometrical units, $8\pi G=1=c$)
\cite{Kanti,John} described by the line element
\begin{eqnarray}
ds^2=-\left(1-\frac{\mu}{r^{n-1}}\right)
dt^2+\left(1-\frac{\mu}{r^{n-1}}\right)^{-1}
dr^2+r^2d\Omega_{n}^{2}
\end{eqnarray}
Here, $r$ being the radial coordinate and
$\mu=\frac{8M~\Gamma\left(\frac{n+1}{2}\right)}{n~\pi^{\frac{n-1}{2}}}$~,
where, $M$ is the mass of the Schwarzschild black hole. Energy
momentum-tensor for the DE, considering in the form of perfect
fluid having the EoS $p=p(\rho)$, is
\begin{eqnarray}
T_{\mu\nu}=(\rho+p)u_\mu u_\nu+p g_{\mu\nu}
\end{eqnarray}
where $\rho$, $p$  are the density and pressure of the dark energy
respectively and $u^\mu=\frac{dx^\mu}{ds}$ is the fluid
$(n+2)$-velocity satisfying $u^\mu u_\mu=-1$. We assume that the
in-falling dark energy fluid does not disturb the spherical symmetry of the black hole.\\

The rate of change of mass $\dot{M}$ of the Schwarzschild black
hole is computed by integrating the flux of the dark energy over
the $n$-dimensional volume of the black hole and given by
\cite{John}
\begin{eqnarray}
\dot{M}=-\frac{2\pi^{\frac{n+1}{2}}}{\Gamma(\frac{n+1}{2})}~r^{n}T_{0}^{1}
\end{eqnarray}
where, $A$ is a positive constant given in \cite{Babichev}, which
can be written as
\begin{eqnarray}
\dot{M}=\frac{2\pi^{\frac{n+1}{2}}}{\Gamma(\frac{n+1}{2})}~AM^{n}(\rho+p)
\end{eqnarray}
For quintessence model, $\rho+p>0$, so $\dot{M}>0$, i.e., $M$
increases. But for phantom model, $\rho+p<0$, so $\dot{M}<0$,
i.e., $M$ decreases as Universe expands.\\

Let us consider $(n+2)$-dimensional spherically symmetrical
accretion of the dark energy onto the wormhole. We consider a
non-static spherically symmetric Morris-Thorne wormhole metric
\cite{Morris} given by
\begin{eqnarray}
ds^2=-e^{\Phi(r)}dt^{2}+\frac{dr^{2}}{1-\frac{K(r)}{r}}
+r^2d\Omega_{n}^{2}
\end{eqnarray}
where the functions $K(r)$ and $\Phi(r)$ are the shape function
and redshift function respectively of radial co-ordinate $r$. If
$K(r_{0})=r_{0}$, the radius $r_{0}$ is called wormhole throat
radius. So we want to consider the outward region such that
$r_{0}\le r<\infty$. Here we have assumed the redshift function
$\Phi(r)=0$. The rate of change of mass $\dot{\bf M}$ of the
wormhole is given by \cite{Diaz4}
\begin{eqnarray}
\dot{\bf
M}=-\frac{2\pi^{\frac{n+1}{2}}}{\Gamma(\frac{n+1}{2})}~B{\bf
M}^{n}(\rho+p)
\end{eqnarray}
where $B$ is positive constant. For quintessence model,
$\rho+p>0$, so $\dot{\bf M}<0$, i.e., ${\bf M}$ decreases. But for
phantom model, $\rho+p<0$, so $\dot{\bf M}>0$, i.e., ${\bf
M}$ increases as Universe expands.\\

We consider the background spacetime is spatially flat represented
by the homogeneous and isotropic $(n+2)$-dimensional FRW model of
the universe which is given by
\begin{eqnarray}
ds^{2}=-dt^{2}+a^{2}(t)\left[dr^{2}+r^{2}d\Omega_{n}^{2}\right]
\end{eqnarray}
where $a(t)$ is the scale factor. The Einstein's field equations
are given by $(8\pi G=c=1)$
\begin{eqnarray}
n(n+1)H^2 = 2 \rho
\end{eqnarray}
and
\begin{eqnarray}
n\dot{H}=-\left(\rho+p \right)
\end{eqnarray}
Conservation equation is
\begin{eqnarray}
\dot{\rho}+(n+1)H(\rho+p)=0
\end{eqnarray}
Now assume that the universe is filled with dark matter and dark
energy, so $\rho=\rho_{m}+\rho_{D}$ and $p=p_{m}+p_{D}$. Here,
$\rho_{m}$ and $p_{m}$ are respectively energy density and
pressure of dark matter. Also, $\rho_{D}$ and $p_{D}$ are
respectively energy density and pressure of dark energy. Now,
assume that the dark matter and dark energy are separately
conserved. So,
\begin{eqnarray}
\dot{\rho}_{m}+(n+1)H(\rho_{m}+p_{m})=0
\end{eqnarray}
and
\begin{eqnarray}
\dot{\rho}_{D}+(n+1)H(\rho_{D}+p_{D})=0
\end{eqnarray}

Now assume that dark matter obeys the equation of state
$p_{m}=w_{m}\rho_{m}$ and using the redshift relation
$1+z=\frac{1}{a}$ (assume, at present, $a_{0}=1$), we get the
solution as
\begin{eqnarray}
\rho_{m}=\rho_{m0}(1+z)^{(1+n)(1+w_{m})}
\end{eqnarray}
where, $\rho_{m0}$ is the present value of the energy density of
dark matter.\\

Using the equations (8) and (10), equation (4) integrates to yield
the mass of the Schwarzschild black hole as
\begin{eqnarray}
M=\frac{M_{0}}{\left[1+\frac{4(n-1)\pi^{\frac{n+1}{2}}AM_{0}^{n-1}}{\Gamma(\frac{n+1}{2})}\sqrt{\frac{n}{2(n+1)}}
~~(\sqrt{\rho}-\sqrt{\rho_{0}})\right]^{\frac{1}{n-1}}}
\end{eqnarray}
where, $\rho_{0}~(=\rho_{m0}+\rho_{D0})$ is the present value of
the density and $M_{0}$ is the present value of the Schwarzschild
black hole mass. Similar happens for wormhole mass. Using equation
(10), equation (6) integrates to yield the mass of the
Morris-Thorne wormhole as in the form:
\begin{eqnarray}
{\bf M}=\frac{{\bf
M}_{0}}{\left[1-\frac{4(n-1)\pi^{\frac{n+1}{2}}B{\bf
M}_{0}^{n-1}}{\Gamma(\frac{n+1}{2})}\sqrt{\frac{n}{2(n+1)}}
~~(\sqrt{\rho}-\sqrt{\rho_{0}})\right]^{\frac{1}{n-1}}}
\end{eqnarray}
where, ${\bf M}_{0}$ is the present value of the Morris-Thorne
wormhole mass. In the following subsections, we shall assume
various types of dark energies like holographic dark energy, new
agegraphic dark energy, quintessence, tachyon, DBI-essence, etc.

\subsection{Holographic Dark Energy}

In quantum field theory a short distance cut-off is related to a
long distance cut-off (infra-red cut-off $L$) due to the limit set
by black hole formation, the total energy in a region of size $L$
should not exceed the mass of a black hole of the same size, i.e.,
$L^{3}\rho_{D}\le LM_{p}^{2}$ (where $M_{p}^{-2}=8\pi G=1$). If
the whole universe is taking into account, then the vacuum energy
related to this holographic principle is viewed as dark energy,
usually called holographic dark energy. The energy density for the
holographic dark energy is given by \cite{Li,Hsu,Cohen}
\begin{equation}
\rho_{D}=3c^{2}L^{-2}
\end{equation}
where $L$ is the IR cut-off length and $c$ is constant. We assume
$L=H^{-1}$. So using equations (8), (13) and (16), we obtain
\begin{equation}
\rho_{D}=\frac{6c^{2}\rho_{m0}}{n(n+1)-6c^{2}}~(1+z)^{(1+n)(1+w_{m})}
\end{equation}
From equation (14), we obtain the mass of the Schwarzschild black
hole as
\begin{eqnarray}
M=\frac{M_{0}}{\left[1+AM_{1}M_{0}^{n-1}~
\left\{(1+z)^{\frac{(1+n)(1+w_{m})}{2}}
-1\right\}\right]^{\frac{1}{n-1}}}
\end{eqnarray}
and from equation (15), we obtain the mass of the wormhole as
\begin{eqnarray}
{\bf M}=\frac{{\bf M}_{0}}{\left[1-BM_{1}{\bf M}_{0}^{n-1}~
\left\{(1+z)^{\frac{(1+n)(1+w_{m})}{2}}
-1\right\}\right]^{\frac{1}{n-1}}}
\end{eqnarray}
where,
$M_{1}=\frac{4n(n-1)\pi^{\frac{n+1}{2}}\sqrt{\rho_{m0}}}{\Gamma(\frac{n+1}{2})\sqrt{2[n(n+1)-6c^{2}]}}$~
with $n(n+1)>6c^{2}$. The black hole mass $M$ and wormhole mass
${\bf M}$ vs redshift $z$ have been drawn in figures 1 and 2
respectively for different values of $n=2,3,4,5$ (i.e.,
$4D,5D,6D,7D$) when dark matter and holographic dark energy
accrete onto black hole and wormhole. For dark matter, we have
taken EoS parameter $w=0.01$. From the figures, we see that black
hole mass increases and wormhole mass decreases during whole
evolution of the Universe. The slope of mass of black hole
increases when $n$ increases i.e., mass of black hole increases
more sharply for increase of dimensions. Similarly, The slope of
mass of wormhole decreases when $n$ increases i.e., mass of
wormhole decreases more sharply for increase of dimensions. These
are the features of dark matter and dark energy accretions onto
black hole and wormhole and natures of increasing/decreasing of
mass completely depends on the dimensions. For more dimension,
mass of the black hole increases more sharply and mass of the
wormhole decreases more sharply for holographic dark energy
accretion.

\begin{figure}
\includegraphics[height=2.0in]{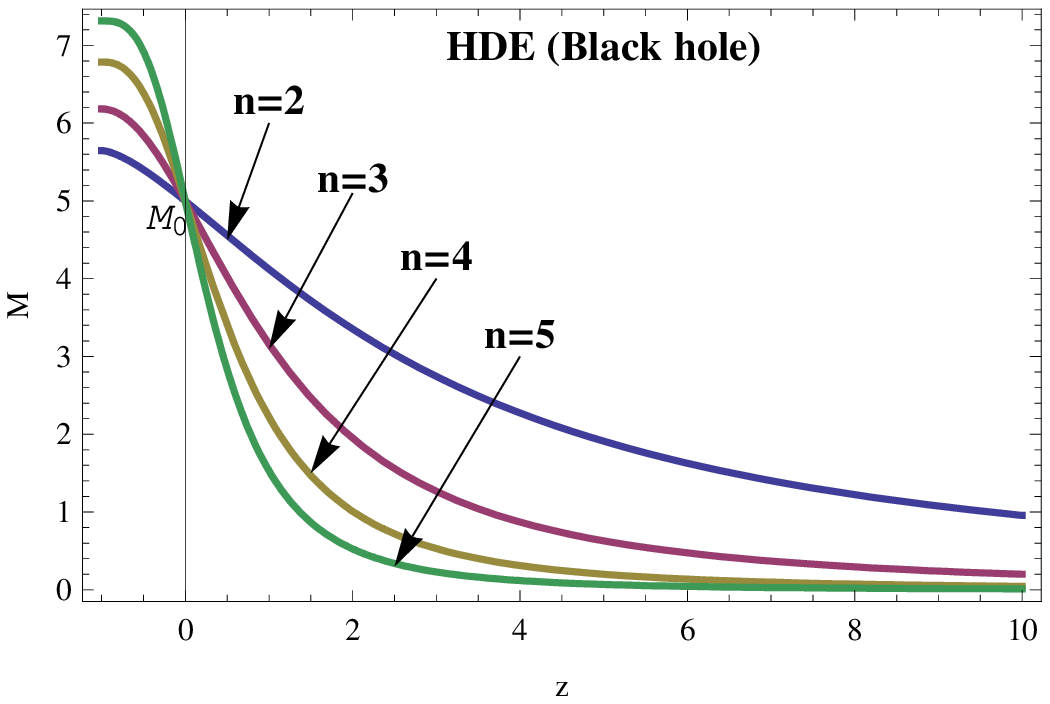}~~~~
\includegraphics[height=2.0in]{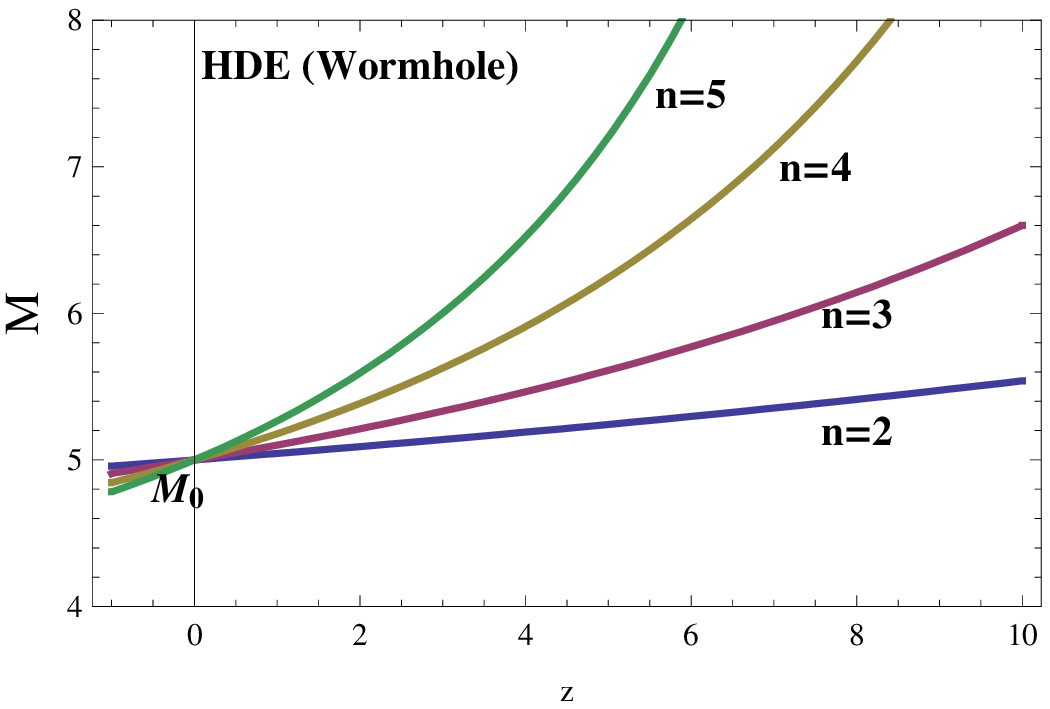}
\vspace{4mm}
~~~~~Fig.1~~~~~~~~~~~~~~~~~~~~~~~~~~~~~~~~~~~~~~~~~~~~~~~~~~~~~~~~~~~~~~~~~~~~~~Fig.2\\
\vspace{4mm} Figs. 1-2 show the variations of black hole mass $M$
and wormhole mass ${\bf M}$ against redshift $z$ respectively for
holographic dark energy accretion in various dimensions
($n=2,3,4,5$). \vspace{0.2in}
\end{figure}

\subsection{New Agegraphic Dark Energy}

The energy density of the new agegraphic dark energy is given by
\cite{Cai01,wei01,wei02}
\begin{equation}
\rho_{D}=\frac{3\alpha^{2}}{\eta^{2}}
\end{equation}
where $\alpha$ is a constant and the conformal time
$\eta=\int_{0}^{t}\frac{dt}{a}$~. For simplicity, we assume the
power law form of the scale factor, $a=a_{i}t^{m}$ where $a_{i}$
is positive constant and $0<m<1$. So we find
$\eta=\frac{t^{1-m}}{a_{i}(1-m)}$~. From above we obtain,
\begin{equation}
\rho_{D}=3\alpha^{2}a_{i}^{2}(1-m)^{2}\left[a_{i}(1+z)\right]^{\frac{2(1-m)}{m}}
\end{equation}
From equation (14), we obtain the mass of the Schwarzschild black
hole as
\begin{eqnarray}
M=\frac{M_{0}}{\left[1+AM_{2}M_{0}^{n-1}~ \left[\{
\sqrt{3\alpha^{2}a_{i}^{2}(1-m)^{2}\left[a_{i}(1+z)\right]^{\frac{2(1-m)}{m}}
+\rho_{m0}(1+z)^{(1+n)(1+w_{m})}}
-\sqrt{\rho_{0}}\right\}\right]^{\frac{1}{n-1}}}
\end{eqnarray}
and from equation (15), we obtain the mass of the wormhole as
\begin{eqnarray}
{\bf M}=\frac{{\bf M}_{0}}{\left[1-BM_{2}{\bf M}_{0}^{n-1}~
\left\{
\sqrt{3\alpha^{2}a_{i}^{2}(1-m)^{2}\left[a_{i}(1+z)\right]^{\frac{2(1-m)}{m}}
+\rho_{m0}(1+z)^{(1+n)(1+w_{m})}}
-\sqrt{\rho_{0}}\right\}\right]^{\frac{1}{n-1}}}
\end{eqnarray}
where,
$\rho_{0}=3\alpha^{2}(1-m)^{2}a_{i}^{\frac{2}{m}}+\rho_{m0}$~~
and~~
$M_{2}=\frac{4(n-1)\pi^{\frac{n+1}{2}}}{\Gamma(\frac{n+1}{2})}\sqrt{\frac{n}{2(n+1)}}$~~.

The black hole mass $M$ and wormhole mass ${\bf M}$ vs redshift
$z$ have been drawn in figures 3 and 4 respectively for different
values of $n=2,3,4,5$ (i.e., $4D,5D,6D,7D$) when dark matter and
new agegraphic dark energy accrete onto black hole and wormhole.
For dark matter, we have taken EoS parameter $w=0.01$. From the
figures, we see that black hole mass increases and wormhole mass
decreases during whole evolution of the Universe. The slope of
mass of black hole increases when $n$ increases i.e., mass of
black hole increases more sharply for increase of dimensions.
Similarly, The slope of mass of wormhole decreases when $n$
increases i.e., mass of wormhole decreases more sharply for
increase of dimensions. These are the features of dark matter and
dark energy accretions onto black hole and wormhole and natures of
increasing/decreasing of mass completely depends on the
dimensions. For more dimension, mass of the black hole increases
more sharply and mass of the wormhole decreases more sharply for
new agegraphic dark energy accretion.

\begin{figure}
\includegraphics[height=2.0in]{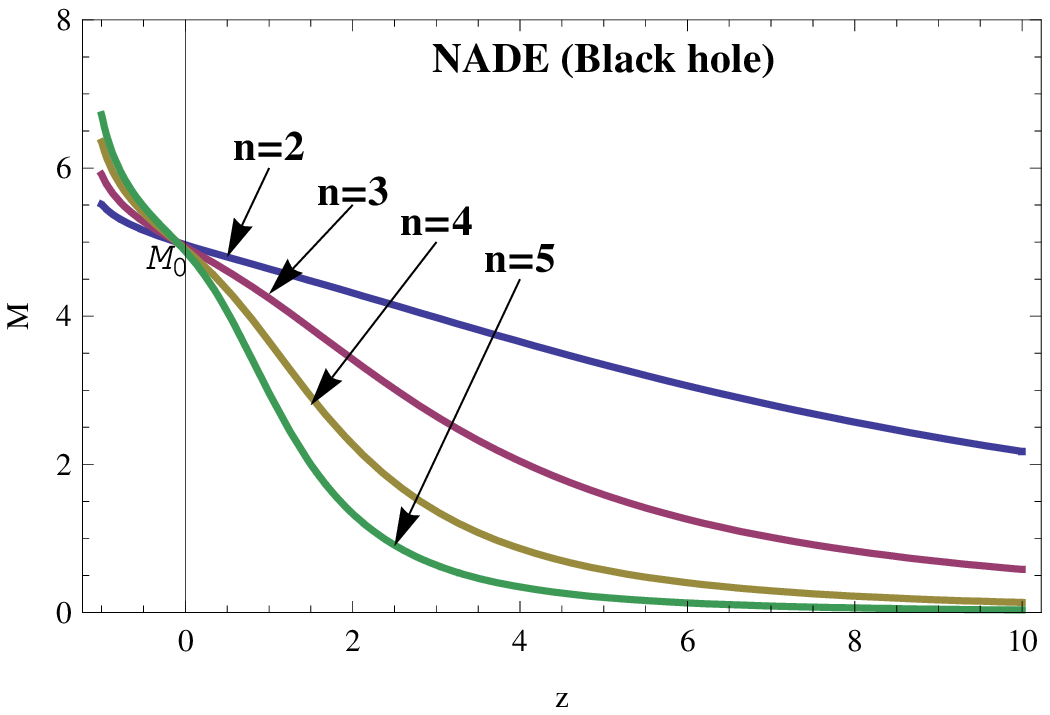}~~~~
\includegraphics[height=2.0in]{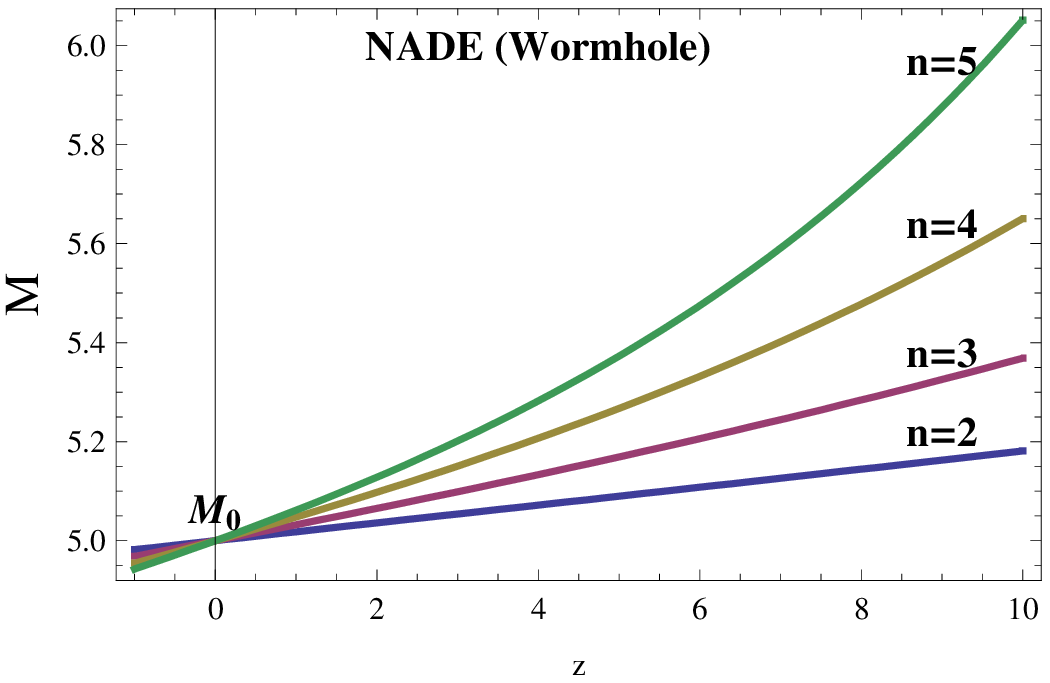}
\vspace{4mm}
~~~~~Fig.3~~~~~~~~~~~~~~~~~~~~~~~~~~~~~~~~~~~~~~~~~~~~~~~~~~~~~~~~~~~~~~~~~~~~~~Fig.4\\
\vspace{4mm} Figs. 3-4 show the variations of black hole mass $M$
and wormhole mass ${\bf M}$ against redshift $z$ respectively for
new agegraphic dark energy (NADE) accretion in various dimensions
($n=2,3,4,5$). \vspace{0.2in}
\end{figure}

\subsection{Quintessence}

The energy density and pressure for quintessence scalar field are
\cite{Peebles,Cald}
\begin{equation}
\rho_{D}=\frac{1}{2}~\dot{\phi}^{2}+V(\phi)
\end{equation}
and
\begin{equation}
p_{D}=\frac{1}{2}~\dot{\phi}^{2}-V(\phi)
\end{equation}
where $\phi$ is the quintessence scalar field and $V(\phi)$ is the
potential. If we put the above expressions in the conservation
equation (12), we get the wave equation, which contains $V$ and
$\dot{\phi}^{2}$. Now to get the solution of $\dot{\phi}^{2}$ and
$V$, we need to consider $V$ in term of either $\phi$ or
$\dot{\phi}^{2}$. If we assume $V$ is some form of $\phi$, it is
very difficult to obtain the solution of $V$ or $\dot{\phi}^{2}$.
So for simplicity of the calculation, we assume the potential term
$V$ is proportional to the kinetic term $\dot{\phi}^{2}$, i.e.,
$V=k\dot{\phi}^{2}$, and we obtain
\begin{equation}
\rho_{D}=\left(k+\frac{1}{2}\right)\left(C(1+z)^{1+n}\right)^{\frac{2}{1+2k}}
\end{equation}
From equation (14), we obtain the mass of the Schwarzschild black
hole as
\begin{eqnarray}
M=\frac{M_{0}}{\left[1+AM_{2}M_{0}^{n-1}~ \left\{
\sqrt{\left(k+\frac{1}{2}\right)
\left\{C(1+z)^{1+n}\right\}^{\frac{2}{1+2k}}
+\rho_{m0}(1+z)^{(1+n)(1+w_{m})}}
-\sqrt{\rho_{0}}\right\}\right]^{\frac{1}{n-1}}}
\end{eqnarray}
and from equation (15), we obtain the mass of the wormhole as
\begin{eqnarray}
{\bf M}=\frac{{\bf M}_{0}}{\left[1-BM_{2}{\bf M}_{0}^{n-1}~
\left\{ \sqrt{\left(k+\frac{1}{2}\right)
\left\{C(1+z)^{1+n}\right\}^{\frac{2}{1+2k}}
+\rho_{m0}(1+z)^{(1+n)(1+w_{m})}}
-\sqrt{\rho_{0}}\right\}\right]^{\frac{1}{n-1}}}
\end{eqnarray} where,
$\rho_{0}=\left(k+\frac{1}{2}\right)C^{\frac{2}{1+2k}}+\rho_{m0}$~~
and~~
$M_{2}=\frac{4(n-1)\pi^{\frac{n+1}{2}}}{\Gamma(\frac{n+1}{2})}\sqrt{\frac{n}{2(n+1)}}$~~.

The black hole mass $M$ and wormhole mass ${\bf M}$ vs redshift
$z$ have been drawn in figures 5 and 6 respectively for different
values of $n=2,3,4,5$ (i.e., $4D,5D,6D,7D$) when dark matter and
quintessence dark energy accrete onto black hole and wormhole. For
dark matter, we have taken EoS parameter $w=0.01$. From the
figures, we see that black hole mass increases and wormhole mass
decreases during whole evolution of the Universe. The slope of
mass of black hole increases when $n$ increases i.e., mass of
black hole increases more sharply for increase of dimensions.
Similarly, The slope of mass of wormhole decreases when $n$
increases i.e., mass of wormhole decreases more sharply for
increase of dimensions. These are the features of dark matter and
dark energy accretions onto black hole and wormhole and natures of
increasing/decreasing of mass completely depends on the
dimensions. For more dimension, mass of the black hole increases
more sharply and mass of the wormhole decreases more sharply for
quintessence dark energy accretion.

\begin{figure}
\includegraphics[height=2.0in]{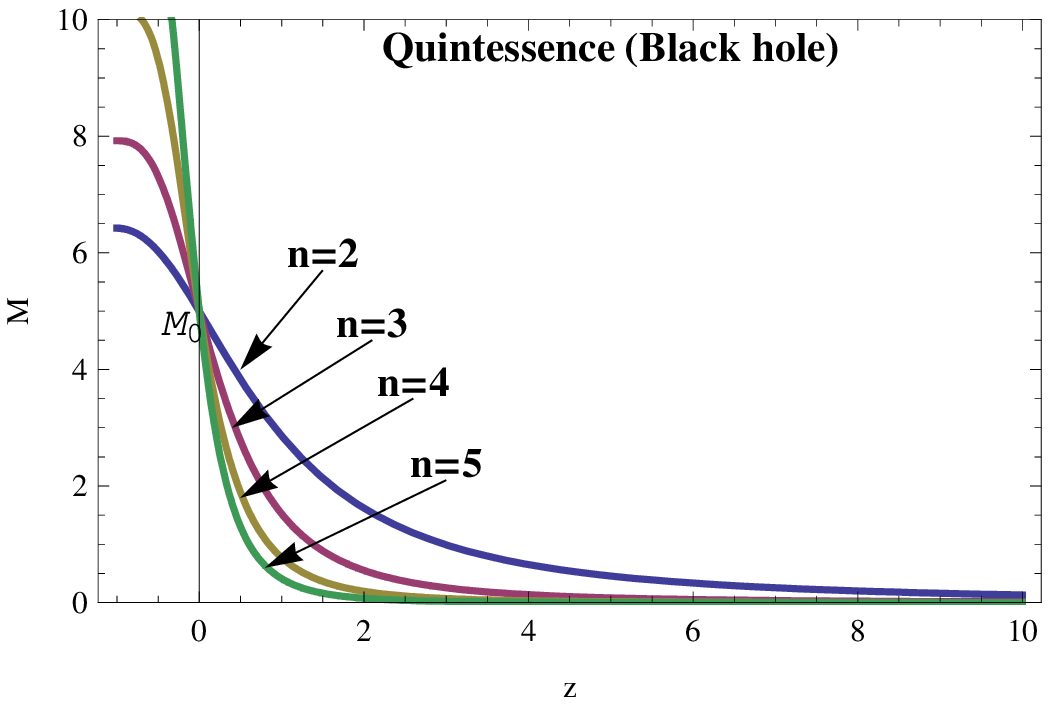}~~~~
\includegraphics[height=2.0in]{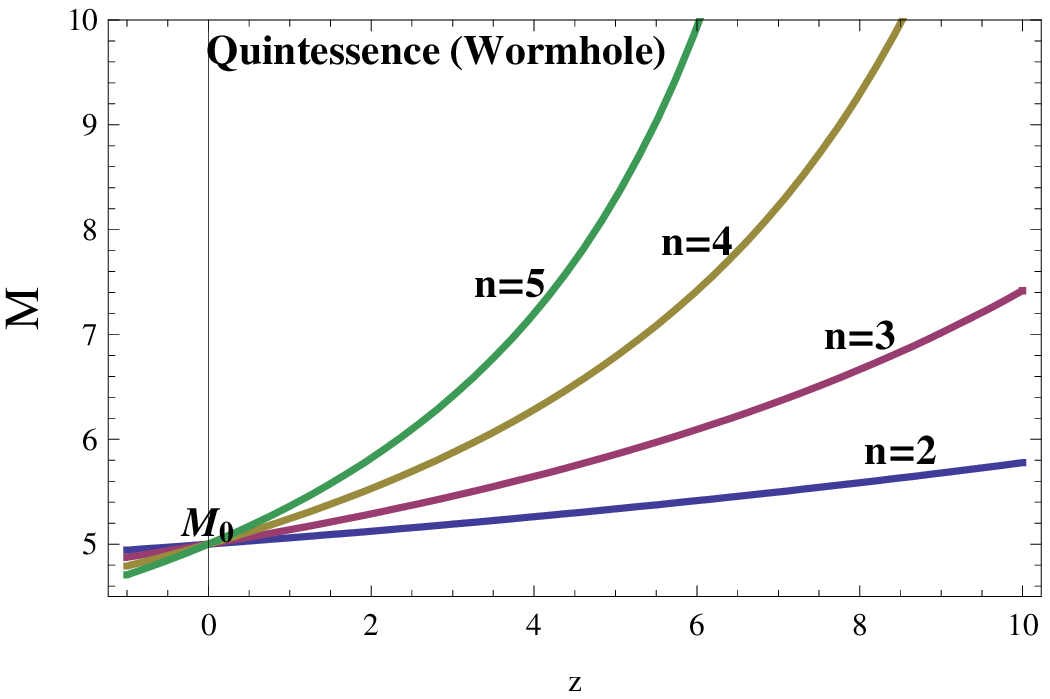}
\vspace{4mm}
~~~~~Fig.5~~~~~~~~~~~~~~~~~~~~~~~~~~~~~~~~~~~~~~~~~~~~~~~~~~~~~~~~~~~~~~~~~~~~~~Fig.6\\
\vspace{4mm} Figs. 5-6 show the variations of black hole mass $M$
and wormhole mass ${\bf M}$ against redshift $z$ respectively for
quintessence accretion in various dimensions ($n=2,3,4,5$).
\vspace{0.2in}
\end{figure}

\subsection{Tachyonic Field}

The energy density $\rho_{D}$ and the pressure $p_{D}$ of the
tachyonic field are \cite{Sen}
\begin{equation}
\rho_{D}=\frac{V(\phi)}{\sqrt{1-{\dot{\phi}}^{2}}}
\end{equation}
and
\begin{equation}
p_{D}=-V(\phi) \sqrt{1-{\dot{\phi}}^{2}}
\end{equation}
where $\phi$ is the tachyonic field, $V(\phi)$ is the tachyonic
field potential. Put these expressions in the conservation
equation (12), we get the wave equation
\begin{equation}
\frac{\dot{V}}{V\dot{\phi}^{2}}+\frac{\ddot{\phi}}{\dot{\phi}}
\left(1-\dot{\phi}^{2}\right)^{-1}+(n+1)H=0
\end{equation}

Now, in order to solve the equation (31), we take a simple form of
$V=\left(1-\dot{\phi}^{2}\right)^{-m}, ~(m>0)$ \cite{Sur}, so that
the solution of $V$ is obtained as
\begin{equation}
V=\left[1+ \left\{C(1+z)^{1+n}\right\}^{\frac{2}{1+2m}}
\right]^{m}
\end{equation}
So from equation (29), we obtain
\begin{equation}
\rho_{D}=\left[1+ \left\{C(1+z)^{1+n}\right\}^{\frac{2}{1+2m}}
\right]^{\frac{1+2m}{2}}
\end{equation}
From equation (14), we obtain the mass of the Schwarzschild black
hole as
\begin{eqnarray}
M=\frac{M_{0}}{\left[1+AM_{2}M_{0}^{n-1}~ \left\{ \sqrt{\left[1+
\left\{C(1+z)^{1+n}\right\}^{\frac{2}{1+2m}}
\right]^{\frac{1+2m}{2}}+\rho_{m0}(1+z)^{(1+n)(1+w_{m})}}
-\sqrt{\rho_{0}}\right\}\right]^{\frac{1}{n-1}}}
\end{eqnarray}
and from equation (15), we obtain the mass of the wormhole as
\begin{eqnarray}
{\bf M}=\frac{{\bf M}_{0}}{\left[1-BM_{2}{\bf M}_{0}^{n-1}~
\left\{ \sqrt{\left[1+
\left\{C(1+z)^{1+n}\right\}^{\frac{2}{1+2m}}
\right]^{\frac{1+2m}{2}}+\rho_{m0}(1+z)^{(1+n)(1+w_{m})}}
-\sqrt{\rho_{0}}\right\}\right]^{\frac{1}{n-1}}}
\end{eqnarray} where,
$\rho_{0}=\left(1+C^{\frac{2}{1+2m}}\right)^{\frac{1+2m}{2}}+\rho_{m0}$~~
and~~
$M_{2}=\frac{4(n-1)\pi^{\frac{n+1}{2}}}{\Gamma(\frac{n+1}{2})}\sqrt{\frac{n}{2(n+1)}}$~~.

The black hole mass $M$ and wormhole mass ${\bf M}$ vs redshift
$z$ have been drawn in figures 7 and 8 respectively for different
values of $n=2,3,4,5$ (i.e., $4D,5D,6D,7D$) when dark matter and
tachyon dark energy accrete onto black hole and wormhole. For dark
matter, we have taken EoS parameter $w=0.01$. From the figures, we
see that black hole mass increases and wormhole mass decreases
during whole evolution of the Universe. The slope of mass of black
hole increases when $n$ increases i.e., mass of black hole
increases more sharply for increase of dimensions. Similarly, The
slope of mass of wormhole decreases when $n$ increases i.e., mass
of wormhole decreases more sharply for increase of dimensions.
These are the features of dark matter and dark energy accretions
onto black hole and wormhole and natures of increasing/decreasing
of mass completely depends on the dimensions. For more dimension,
mass of the black hole increases more sharply and mass of the
wormhole decreases more sharply for tachyon dark energy accretion.

\begin{figure}
\includegraphics[height=2.0in]{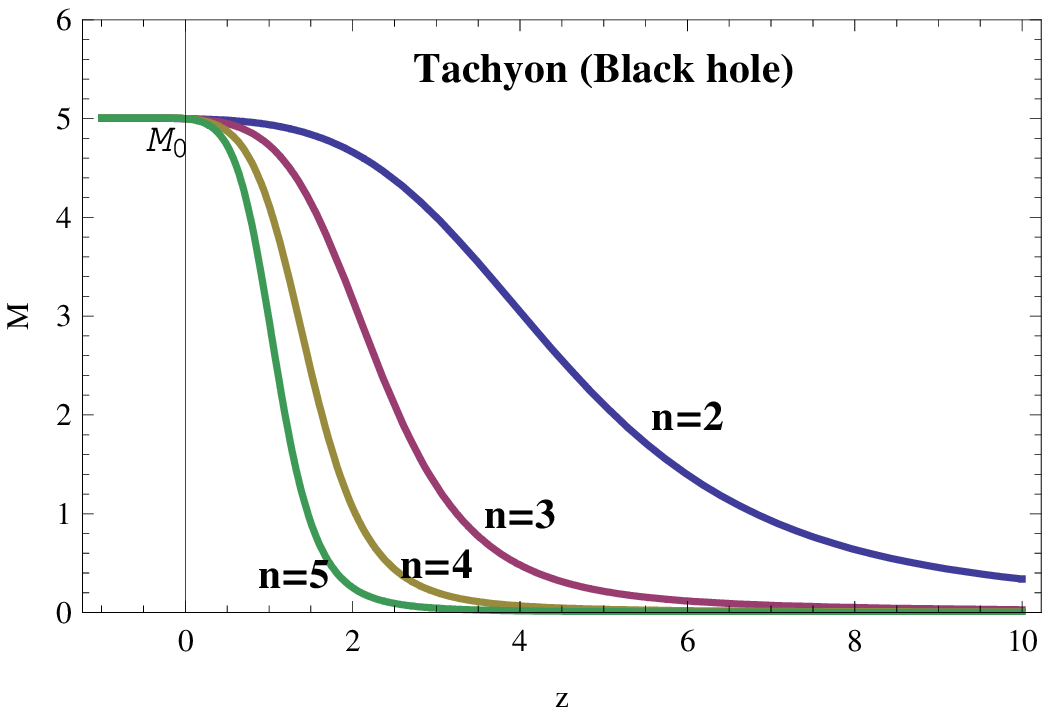}~~~~
\includegraphics[height=2.0in]{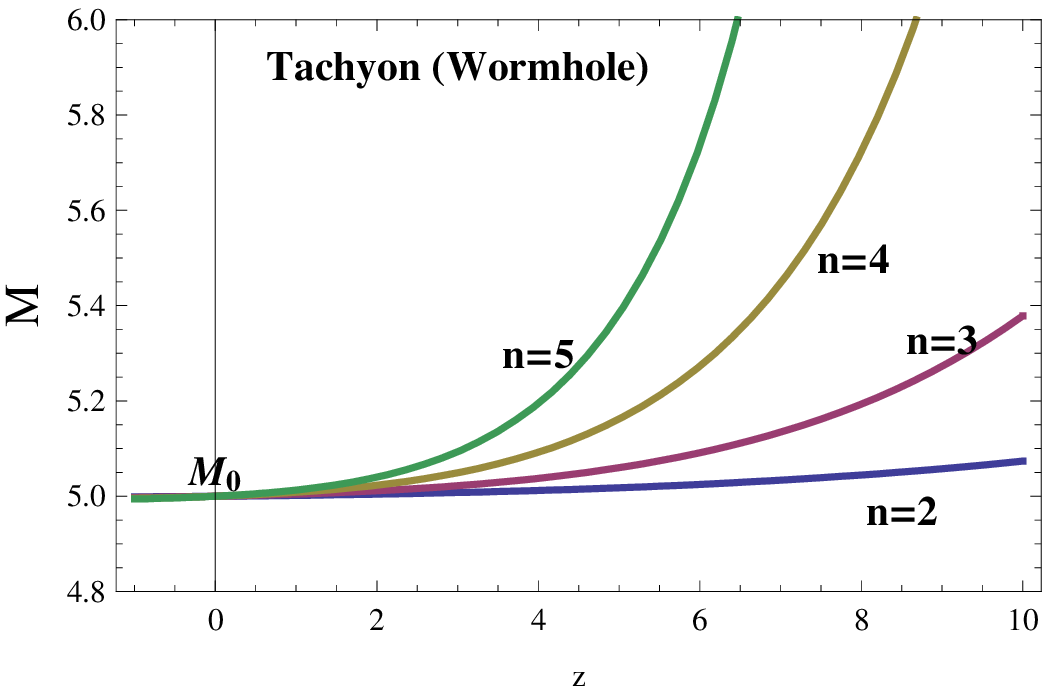}
\vspace{4mm}
~~~~~Fig.7~~~~~~~~~~~~~~~~~~~~~~~~~~~~~~~~~~~~~~~~~~~~~~~~~~~~~~~~~~~~~~~~~~~~~~Fig.8\\
\vspace{4mm} Figs. 7-8 show the variations of black hole mass $M$
and wormhole mass ${\bf M}$ against redshift $z$ respectively for
Tachyonic field accretion in various dimensions ($n=2,3,4,5$).
\vspace{0.2in}
\end{figure}

\subsection{DBI-essence}

The energy density and pressure of the scalar field are
respectively given by \cite{Gum,Mart}
\begin{equation}
\rho_{D}=(\gamma-1)T(\phi)+V(\phi),
\end{equation}
\begin{equation}
p_{D}=\frac{\gamma-1}{\gamma}~T(\phi)-V(\phi),
\end{equation}
where the quantity $\gamma$ is reminiscent from the usual
relativistic Lorentz factor and is given by
\begin{equation}
\gamma=\frac{1}{\sqrt{1-\frac{\dot{\phi}^{2}}{T(\phi)}}}.
\end{equation}
where $\phi$ is the DBI scalar field and $V(\phi)$ is the
corresponding potential. From energy conservation equation (12),
we have the wave equation for $\phi$ as
\begin{equation}
\ddot{\phi}-\frac{3T'(\phi)}{2T(\phi)}~\dot{\phi}^{2}+
T'(\phi)+\frac{(n+1)}{\gamma^{2}}~\frac{\dot{a}}{a}~\dot{\phi}+
\frac{1}{\gamma^{3}}[V'(\phi)-T'(\phi)]=0.
\end{equation}
Let us assume, $\gamma=\dot{\phi}^{-2}$~\cite{J1}, so from (38) we
have $T(\phi)=\frac{\dot{\phi}^{2}}{1-\dot{\phi}^{4}}$. Let us
also assume $V(\phi)=T(\phi)$. In this case, we have the
solutions:
\begin{equation}
\dot{\phi}^{2}=\sqrt{1+\frac{1}{(n+1)\log\frac{C}{a}}}
\end{equation}
\begin{equation}
V(\phi)=T(\phi)=(n+1)\log\frac{a}{C}\times\sqrt{1+\frac{1}{(n+1)\log\frac{C}{a}}}
\end{equation}
where $C$ is the integration constant. Thus from equation (36), we
obtain
\begin{equation}
\rho_{D}=(n+1)\log\frac{1}{C(1+z)}
\end{equation}
From equation (14), we obtain the mass of the Schwarzschild black
hole as
\begin{eqnarray}
M=\frac{M_{0}}{\left[1+AM_{2}M_{0}^{n-1}~ \left\{
\sqrt{(n+1)\log\frac{1}{C(1+z)}+\rho_{m0}(1+z)^{(1+n)(1+w_{m})}}
-\sqrt{\rho_{0}}\right\}\right]^{\frac{1}{n-1}}}
\end{eqnarray}
and from equation (15), we obtain the mass of the wormhole as
\begin{eqnarray}
{\bf M}=\frac{{\bf M}_{0}}{\left[1-BM_{2}{\bf M}_{0}^{n-1}~
\left\{
\sqrt{(n+1)\log\frac{1}{C(1+z)}+\rho_{m0}(1+z)^{(1+n)(1+w_{m})}}
-\sqrt{\rho_{0}}\right\}\right]^{\frac{1}{n-1}}}
\end{eqnarray} where, $\rho_{0}=(n+1)\log\frac{1}{C}+\rho_{m0}$~~ and~~
$M_{2}=\frac{4(n-1)\pi^{\frac{n+1}{2}}}{\Gamma(\frac{n+1}{2})}\sqrt{\frac{n}{2(n+1)}}$~~.

The black hole mass $M$ and wormhole mass ${\bf M}$ vs redshift
$z$ have been drawn in figures 9 and 10 respectively for different
values of $n=2,3,4,5$ (i.e., $4D,5D,6D,7D$) when dark matter and
DBI-essence accrete onto black hole and wormhole. For dark matter,
we have taken EoS parameter $w=0.01$. From the figures, we see
that black hole mass first increases upto a certain finite value
and then decreases and wormhole mass decreases upto a certain
finite value and then increases during whole evolution of the
Universe. These are the features of dark matter and dark energy
accretions onto black hole and wormhole for DBI-essence accretion,
because our considered model is the phantom crossing model. The
natures of increasing/decreasing of mass are nearly similar to all
dimensions.

\begin{figure}
\includegraphics[height=2.0in]{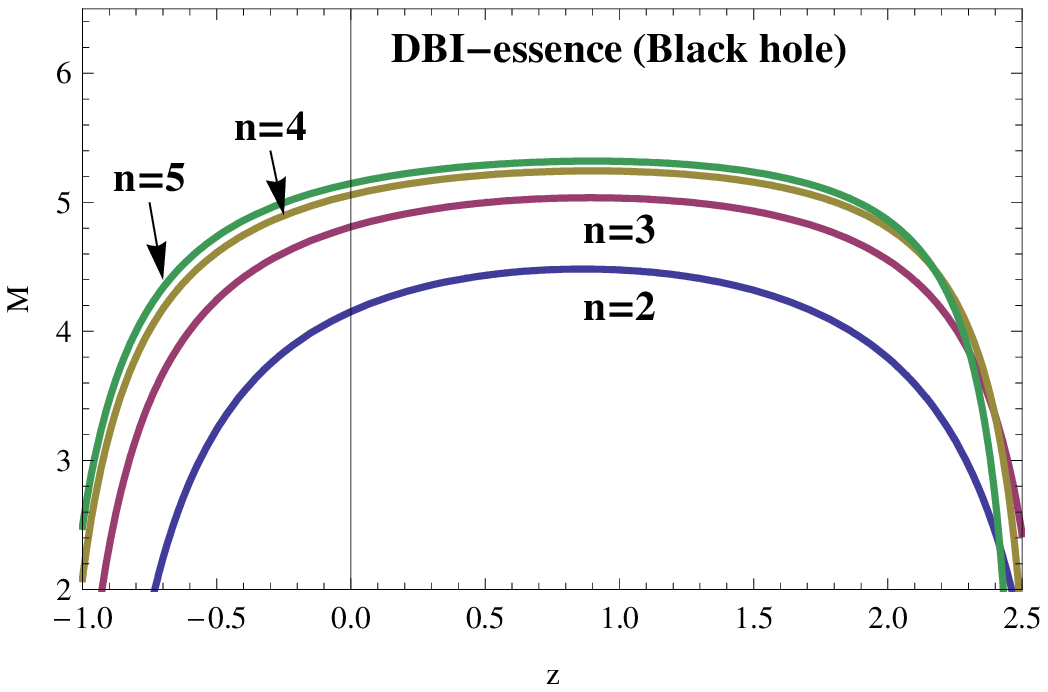}~~~~
\includegraphics[height=2.0in]{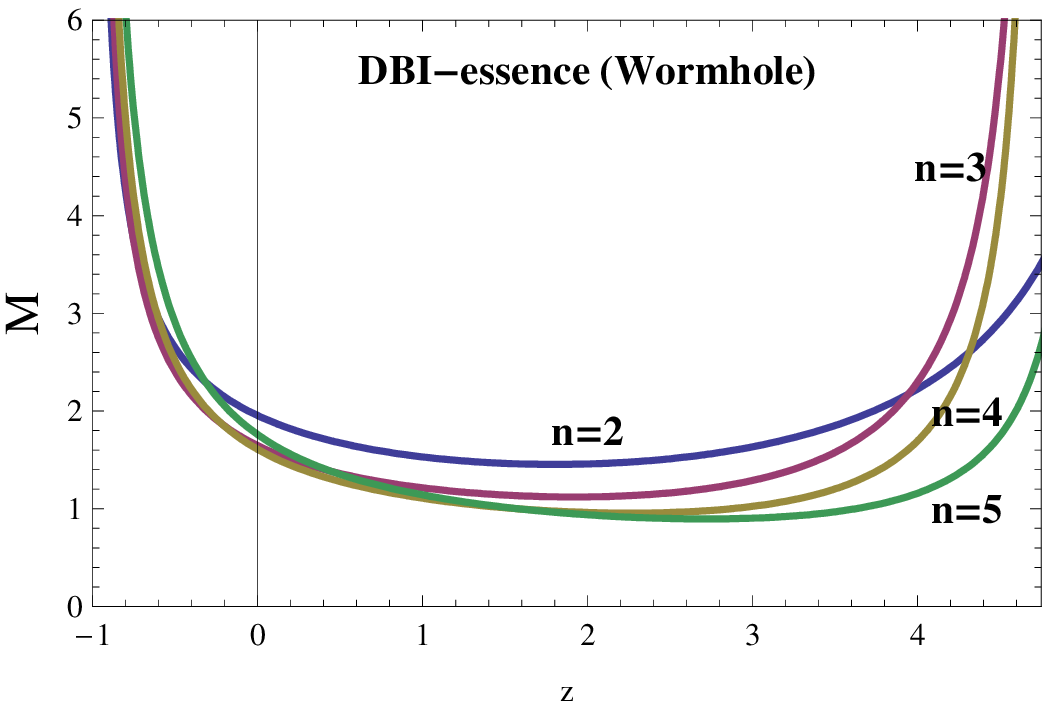}
\vspace{4mm}
~~~~~Fig.9~~~~~~~~~~~~~~~~~~~~~~~~~~~~~~~~~~~~~~~~~~~~~~~~~~~~~~~~~~~~~~~~~~~~~~Fig.10\\
\vspace{4mm} Figs. 9-10 show the variations of black hole mass $M$
and wormhole mass ${\bf M}$ against redshift $z$ respectively for
DBI-essence accretion in various dimensions ($n=2,3,4,5$).
\vspace{0.2in}
\end{figure}

\section{Discussions and Concluding Remarks}

A proper dark-energy accretion model for black hole have been
obtained by generalizing the Michel theory \cite{Michel} to the
case of black hole. Such a generalization has been already
performed by Babichev et al \cite{Babichev,Babichev1} for the case
of dark-energy accretion onto Schwarzschild black hole.
Astrophysically, masses of the black hole and wormhole are
dynamical quantity, so the nature of the mass function is
important in our black hole/wormhole model for different dark
energy filled universe. Previously it has shown that the mass of
black hole increases due to quintessence energy accretion and
decreases due to phantom energy accretion. In ref \cite{Diaz4}, it
was shown that for quintessence like dark energy, the mass of the
wormhole decreases and phantom like dark energy, the mass of
wormhole increases, which is the opposite behaviour of black hole
mass.\\

In this work, we have studied accretion of the dark matter and
dark energy onto of $(n+2)$-dimensional Schwarzschild black hole
and Morris-Thorne wormhole. The mass and the rate of change of
mass for $(n+2)$-dimensional Schwarzschild black hole and
Morris-Thorne wormhole have been found. We have assumed some
candidates of dark energy like holographic dark energy, new
agegraphic dark energy, quintessence, tachyon, DBI-essence, etc.
The black hole mass and the wormhole mass have been calculated in
term of redshift when dark matter and above types of dark energies
accrete onto them separately. The black hole mass $M$ vs redshift
$z$ have been drawn in figures 1, 3, 5, 7 respectively for
different values of $n=2,3,4,5$ (i.e., $4D,5D,6D,7D$) when dark
matter and holographic dark energy, agegraphic dark energy,
quintessence, tachyon accrete onto black hole. For dark matter, we
have taken EoS parameter $w=0.01$. From the figures, we see that
black hole mass increases during whole evolution of the Universe.
The slope of mass of black hole increases when $n$ increases i.e.,
mass of black hole increases more sharply for increase of
dimensions. On the other hand, the wormhole mass ${\bf M}$ vs
redshift $z$ have been drawn in figures 2, 4, 6, 8 respectively
for different values of $n=2,3,4,5$ (i.e., $4D,5D,6D,7D$) when
dark matter and holographic dark energy, agegraphic dark energy,
quintessence, tachyon accrete onto wormhole. From the figures, we
see that wormhole mass decreases during whole evolution of the
Universe. The slope of mass of wormhole decreases when $n$
increases i.e., mass of wormhole decreases more sharply for
increase of dimensions. These are the features of dark matter and
dark energy accretions onto black hole and wormhole and natures of
increasing/decreasing of mass completely depends on the
dimensions. For more dimension, mass of the black hole increases
more sharply and mass of the wormhole decreases more sharply for
holographic dark energy, agegraphic dark energy, quintessence,
tachyon accretion.\\

Lastly, the black hole mass $M$ and wormhole mass ${\bf M}$ vs
redshift $z$ have been drawn in figures 9 and 10 respectively for
different values of $n=2,3,4,5$ (i.e., $4D,5D,6D,7D$) when dark
matter and DBI-essence accrete onto black hole and wormhole. From
the figures, we see that black hole mass first increases upto a
certain finite value and then decreases and wormhole mass
decreases upto a certain finite value and then increases during
whole evolution of the Universe. These are the features of dark
matter and dark energy accretions onto black hole and wormhole for
DBI-essence accretion, because our considered model is the phantom
crossing model. The natures of increasing/decreasing of mass are
nearly similar to all dimensions. Hence, we conclude that the
black hole mass increases and wormhole mass decreases for
holographic dark energy, new agegraphic dark energy, quintessence,
tachyon accretion and the slope of increasing/decreasing of mass
sensitively depends on the dimension. But for DBI-essence
accretion, the black hole mass first increases and then decreases
and the wormhole mass first decreases and then increases and the
slope of increasing/decreasing of mass not sensitively depends on
the dimension.\\

{\bf Acknowledgement:}\\

The author is thankful to IUCAA, Pune, India for warm hospitality
where the work was carried out.\\


\begin{thebibliography}{99}
\bibitem{Bachall} N. A. Bachall, J. P. Ostriker, S. Perlmutter and P. J. Steinhardt,
{\it Science} {\bf 284} 1481 (1999).
\bibitem{Perlmutter1} S. J. Perlmutter et al, {\it Bull. Am. Astron. Soc.} {\bf 29}, 1351 (1997).
\bibitem{Perlmutter2} S. J. Perlmutter et al, {\it Astrophys. J.} {\bf 517} 565 (1999).
\bibitem {Riess} A. G. Riess et al, {\it Astron. J.} {\bf 116}, 1009 (1998).
\bibitem{Bern} P. de Bernardis et al, Nature 404, 955 (2000).
\bibitem{Lange} A.E. Lange et al, Phys. Rev. D 63, 042001 (2001).
\bibitem{Babli} A. Balbi et al, Astrophys. J. 545, L1 (2000).
\bibitem{Briddle} Briddle, S. et al, 2003, Science 299, 1532.
\bibitem{Spergel} Spergel, D. N. et al, 2003, Astrophys. J. Suppl. 148,
175.
\bibitem{Peebles} P. J. E. Peebles and B. Ratra, \textit{Astrophys. J.}
\textbf{325} L17 (1988).
\bibitem{Cald} R. R. Caldwell, R. Dave and P. J. Steinhardt, \textit{Phys.
Rev. Lett.} \textbf{80} 1582 (1998).
\bibitem{Bennett} C. L. Bennett et al, {\it Astrophys. J. Suppl.} 148, 1 (2003).
\bibitem{Allen} S. W. Allen et al, {\it Mon. Not. Roy. Astron. Soc.} 353, 457 (2004).
\bibitem{Alam2} U. Alam, V. Sahni, T. D. Saini, and A. A. Starobinsky, Mon.Not.Roy.Astron.Soc. 354, 275 (2004).
\bibitem{Gum} B. Gumjudpai and J. Ward, Phys. Rev. D 80 023528 (2009).
\bibitem{Mart} J. Martin and M. Yamaguchi, Phys. Rev. D 77 123508
(2008).
\bibitem{Sen} A. Sen, \textit{JHEP} \textbf{0207} 065 (2002).
\bibitem{Li}M.Li,Phys.Lett.B 603(2004)1,hep-th/0403127.
\bibitem{Hsu}S.D.H.Hsu ,Phys.Lett.B 594(2004)13,hep-th/0403052.
\bibitem{Cohen}A.G.Cohen,D.B.Kaplan,A.E.Nelson,Phys.Rev.D Lett.82(1999)4971 ,hep-th/98803132.
\bibitem{Cai01} H. Wei, R. -G. Cai, Phys. Lett. B, 660 (2008) 113.
\bibitem{wei01}H.Wei,R.G.Cai,Eur.Phys.J.C59:99,2009.
\bibitem{wei02} H.Wei,R.G.Cai,Phys.Lett.B 655(2007)1.
\bibitem{Bondi} H. Bondi, Mon. Not. Roy. Astron. Soc. 112, 195 (1952).
\bibitem{Michel}F. C. Michel, Astrophys. Space Sci. 15, 153 (1972).
\bibitem{Babichev} E. Babichev et al, 2004 Phys. Rev. Lett. 93, 021102.
\bibitem{Babichev1} E. Babichev, V. Dokuchaev, Y. Eroshenko, J.Exp.Theor.Phys. 100
(2005) 528-538.
\bibitem{Jamil}M. Jamil, Eur.Phys.J.C62:609,2009.
\bibitem{Pedro}Jo$\acute{s}$e A. Ji$\acute{m}$enez Madrid, and
Pedro F. Gon$\acute{z}$alez-D$\acute{i}$az, Grav. Cosmol. 14, 213
(2008).
\bibitem{Bhadra} J. Bhadra and U. Debnath,  Eur. Phys. J. C. 72, 1912 (2012).
\bibitem{Nayak} B. Nayak and M. Jamil, Phys. Lett. B 709, 118 (2012).
\bibitem{Lima} J.A.S. Lima, D. C. Guariento and J.E. Horvath, Phys. Lett. B 693, 218
(2010).
\bibitem{Sharif1} M. Sharif and G. Abbas, Chinese Phys. Lett. 29,
010401 (2012).
\bibitem{Sun} C. Y. Sun, Phys. Rev. D 78, 064060 (2008).
\bibitem{Kim} S. W. Kim and Y. Kang, Int. J. Mod. Phys. Conf. Ser. 12, 320 (2012).
\bibitem{Rod} M. G. Rodrigues and A. E. Bernardiniz, arXiv:1208.1572v1
[gr-qc].
\bibitem{Abhas} G. Abbas, arXiv:1303.6945v1 [gr-qc].
\bibitem{Mar1} P. Mart$\acute{i}$n-Moruno et al, arXiv:astro-ph/0603761.
\bibitem{Abha} G. Abhas, arXiv: 1309.0807 [gr-qc].
\bibitem{Diaz1} P.F. Gonz$\acute{a}$lez-D$\acute{i}$az, Phys. Rev. Lett. 93, 071301 (2004).
\bibitem{Far} V. Faraoni, W. Israel, Phys. Rev. D 71, 064017 (2005).
\bibitem{Diaz2} P. F. Gonz$\acute{a}$lez-D$\acute{i}$az, arXiv:astro-ph/0404045.
\bibitem{Diaz3} P. F. Gonz$\acute{a}$lez-D$\acute{i}$az, Phys. Lett. B 632, 159 (2006).
\bibitem{Mad} J. A. J. Madrid and P. Mart$\acute{i}$n-Moruno,
arXiv:1004.1428 [astro-ph.CO].
\bibitem{Mor} P. Mart$\acute{i}$n-Moruno, Phys. Lett. B 659, 40 (2008).
\bibitem{Diaz4} P.F. Gonz$\acute{a}$lez-D$\acute{i}$az, arXiv:hep-th/0607137.
\bibitem{Diaz5} P.F. Gonz$\acute{a}$lez-D$\acute{i}$az and P. Mart$\acute{i}$n-Moruno, arXiv:0704.1731v1
[astro-ph].
\bibitem{Kim1} S. P. Kim, S. K. Kim, K. S. Soh and J. H. Yee, Phys.
Rev. D 55 (1997) 2159.
\bibitem{Ghosh} S. G. Ghosh, arXiv:1109.3263v2 [gr-qc].
\bibitem{John} A. J. John, S. G. Ghosh, S. D. Maharaj,
Phys. Rev. D 88, 104005 (2013).
\bibitem{Hendi} S. H. Hendi, Eur.Phys.J.C71:1551,2011.
\bibitem{Kanti} P. Kanti and E. Winstanley, arXiv:1402.3952[hep-th].
\bibitem{Morris} M. S. Morris and K. S. Thorne, Am. J. Phys. 56, 395 (1988).
\bibitem{Sur} S. Chattopadhyay, U. Debnath and G. Chattopadhyay, Astrophys. Space Sci.,
 314 (2008) 41.
\bibitem{J1} U. Debnath and M. Jamil, Astrophys. Space Sci. 335,
545 (2011).
\end{thebibliography}
\end{document}